# Getting Virtualized Wireless Sensor Networks' IaaS Ready for PaaS

Imran Khan*†, Fatima Zahra Errounda†, Sami Yangui†, Roch Glitho† and Noel Crespi¨ *

*Institut Mines-Télécom, Télécom SudParis, 91011 Evry Cedex, France
Email: imran@ieee.org, noel.crespi@it-sudparis.eu

†Dept. CIISE, Concordia University, H3G 2W1, Montreal, Canada
Email: {f errou, s yangui} @encs.concordia.ca, glitho@ciise.concordia.ca

*Abstract*—With the recent advances in sensor hardware and software, architectures for virtualized Wireless Sensor Networks (vWSNs) are now emerging. Through node- and network-level virtualization, vWSNs can be offered as Infrastructure-as-a-Service (IaaS) which can aid in realizing the true potential of Internet-of-Things (IoT). Cloud computing offers elastic provisioning of large-scale infrastructures to multiple concurrent users where Platform-as-a-Service (PaaS) interacts with IaaS in order to efficiently host and execute applications over these infrastructures. Amalgamating IoT with cloud computing potentially allows rapid application and service provisioning in an efficient, scalable and robust manner. However, interactions between vWSNs and PaaS are largely an unexplored area. Indeed, existing vWSN IaaS are not yet ready for PaaS. This paper proposes a vWSN IaaS architecture which is ready for interactions with PaaS. The proposed architecture is based on our previous works and is rooted in the fundamental differences between traditional IaaS and vWSN IaaS. We built a prototype using Java Sunspot as the WSN tool kit and made early performance measurements.

*Keywords*—*Wireless Sensor Networks; Internet of Things; Cloud Computing; Virtualization; IaaS; PaaS*

## I. INTRODUCTION

Since their mainstream introduction towards the end of 20th century, Wireless Sensor Network (WSN) deployments have been used as means to bridge the gap between the physical world and the virtual world. With their ability to sense, compute and communicate, WSNs provide their users with the ability to react to various physical phenomenon and take required actions [1]. WSNs are considered as basic building blocks of Internet-of-Things (IoT) paradigm [2] where sensors, along with multitude of everyday objects communicate, interact and share data on a massive scale [3].

Cloud computing [4] paradigm allows several inherent benefits (e.g., efficient usage of resources, scalability, elasticity, and rapid provisioning of new applications). It has three key facets: Software-as-a-Service (SaaS), Platform-as-aService (PaaS) and Infrastructure-as-a-Service (IaaS). Service providers use PaaS to provision applications and services as SaaS on a pay-per-use basis to the end-users. PaaS ease the provisioning process by adding levels of abstraction to the infrastructure. This abstraction is achieved by using the virtualization concept that allows sharing of resources by abstracting them into multiple logical units on the same physical node [5].

WSNs can be virtualized at node-level [6] as well as at network-level [7]. At node-level, multiple applications can run tasks concurrently on a single WSN node, either sequentially (round-robin) or simultaneously (context switching). At network-level, groups of WSN nodes form Virtual Sensor Networks (VSNs) to execute a given application task at a given time. There can be multiple such groups in a WSN deployment, each dedicated to a different application. A detailed survey discussing the basics, motivation, benefits and existing works on WSN virtualization can be found in [8].

Architectures that combine WSN node- and network-level virtualization are now emerging (e.g., [9], [10] and [11]). However, they are still not yet ready for PaaS. They lack the appropriate design and architectural details to enable proper interactions with the PaaS so that service providers are able to efficiently provision new WSN applications and services. The problem is challenging because vWSN IaaS are fundamentally different from traditional IaaS. For example, in traditional IaaS the concept of Virtual Machine (VM) is used, which is characterized by its operating system, unique global address, processing power and memory. On the other hand, in vWSNs the concept of Virtual Sensor (VS) is used, which is characterized by its sensor middleware, platformdependent localized address and scarcity of processing power and memory. Moreover, issues like geospatial location and sampling rate impose additional constraints.

This paper proposes an architecture to offer competent vWSN IaaS, which is able to interact with PaaS to allow service providers to rapidly provision WSN-based applications and services. The proposed architecture is based on our previous work [10] and on the fundamental differences between the vWSN IaaS and traditional IaaS that we have identified. Unlike our previous work, this paper focuses on architectural design and details to enable interactions between vWSN IaaS and PaaS for dynamic provisioning of applications and services.

The paper is organized as follows. In Section II differences between traditional IaaS and vWSN IaaS are presented along with the requirements for a PaaS ready vWSN IaaS. Section III presents the proposed vWSN architecture. Details on the implementation and the results are presented in Section VI. Section V discusses the lessons learned and future work while Section VI concludes the paper.

## II. FUNDAMENTAL DIFFERENCES BETWEEN WSN IaaS AND TRADITIONAL IaaS

The fundamental differences between vWSN IaaS and traditional IaaS stem from the differences between WSNs and traditional networks. In this section, we first briefly discuss how WSN and traditional networks differ before introducing the fundamental differences between vWSN IaaS and traditional IaaS. Our analysis will be structured around the concepts of VM (i.e., the fundamental element of traditional IaaS) and VS (i.e., the fundamental element of vWSN IaaS). Finally, we present a set of requirements for a PaaS ready vWSN IaaS.

### A. Differences between WSN and Traditional Networks

WSNs are known to be resource-constrained environments whose nodes typically have limited processing capability, storage and are battery operated. The nodes have low duty cycle [12] and operate only at specific intervals [13]. This means that WSN nodes are not always available for applications. In traditional networks, nodes (server, computers) have considerable resources and potentially have unlimited power source allowing high duty cycle and high availability. This fundamental difference has led to numerous research efforts aimed at designing energy efficient protocols [14], simple data formats [15] and simple application design [16] for WSNs. Another important difference between the two network types is the availability of protocols. IP rules traditional networks whereas in WSN it is not much prevalent yet but there have been efforts to bring IP to the WSN world [17], [18] and [19]. HTTP is not as much useful in WSNs as in traditional networks but alternatives like CoAP [20] have emerged for WSNs. We observe that the advent of IoT paradigm has prompted many efforts to provide standard protocol support for WSNs [21].

### B. Differences between VM and VS

A VM is defined as a logical unit that allows time and resource sharing of host machines by partitioning them into multiple dedicated execution environments [22]. Each VM has a guest operating system that can access underlying resources. On the other hand, a VS is a logical representation of the physical sensor to allow sharing of its sensing capabilities (e.g., temperature and light sensing capabilities) [10]. VSs execute multiple concurrent application tasks. On an abstract level, a VS is similar to a VM, i.e., both provide a mechanism to decouple physical resources from their host nodes in order to be used by multiple users. For example, in traditional IaaS, the resources of a host machine are represented by a VM Monitor (VMM) or Hypervisor that allows multiple VMs to access underlying resources [23]. In vWSN IaaS, if we consider the example of Java SunSpots, the Squawk virtual machine [24] provides a similar type of abstraction that allows multiple VSs to access the sensing resources of a sensor. Still, there are certain fundamental differences between the two. Table I lists seven such differences, which are explained below.

The *first* difference is that a VM allows for the sharing of resources (e.g., computing and storage) of the host machine, whereas a VS allows sharing of sensing capabilities (e.g., temperature, light, humidity) by executing multiple application tasks. The key difference is that a VM aims at sharing the host machine resources, whereas a VS may use the computing and storage of the host sensor, but it aims at sharing the sensing capabilities of the host sensor. In Java SunSpots, for instance, application tasks access the on-board sensors to sense the physical phenomenon, and send the data accordingly.

The *second* difference is that multiple heterogeneous VMs (in terms of operating systems) can be simultaneously deployed on the same host. For instance, a host can support a Linux-based VM and/or a Windows-based VM at the same time. However, VSs are tightly coupled with their sensor OS/middleware. For example, a sensor cannot support Contikibased VS and TinyOS-based VS at the same time.

The *third* difference is that multiple VMs can be deployed in an isolated manner. The creation, deployment, and migration of VMs does not affect the execution of existing VMs. On the other hand, the deployment of new VS may disturb the execution of existing VS(s). This is due to the limited resources and the tight coupling between the VS and the sensor OS/middleware. Similarly, migrating VS from one physical sensor to another is not a standard feature yet. To the best of our knowledge, Java SunSpots is the only platform that provides support for VS migration (as serialized Java Isolates). There is additional work in which a mobile agent-based system for Java SunSpots is developed for VS migration [25].

The *fourth* difference is that VMs can be addressed by other entities that are similar to their host machines. Each VM can be assigned a public or private IP address and can be accessed accordingly. However, there is currently no standard mechanism for addressing a VS. Typically, a local ID is used and may vary depending on the platform. This necessitates some address mapping/translation mechanism to communicate with a VS. For instance, in Java SunSpots, each VS can be addressed by a MIDlet ID.

The *fifth* difference is that for a VM, there are no power/energy-related issues, whereas a VS inherits these issues from the host sensor nodes. This means that the creation, deployment, and operation of a VS are not only dependent on the capabilities/resources of the host sensor, but also on its available energy. The always-on/always-available concept is not applicable to WSN world.

The *sixth* difference is that for VMs, there are already some open source and proprietary solutions (e.g., KVM and VMware). However, no such solutions exist for VSs.

The *seventh and final* difference is that, at the IaaS level, the role of a VM is to maximize the use of a host machines resources (e.g., computing and storage), while the role of a VS is to use the sensing capabilities of the host sensor in an efficient manner. Therefore, to achieve cost-efficiency, traditional IaaS may create several VMs on a limited number of host machines. However, achieving cost-efficiency in vWSN IaaS may not lead to the creation of several VSs on a few host sensor nodes since the deployment of sensor nodes is strongly correlated to the desired coverage of a geographic area.

### C. Requirements for a PaaS Ready vWSN IaaS

The *first* requirement is that vWSN IaaS should support standard interfaces for interacting with a PaaS API. These standardized interfaces will allow easy instantiation, operation and management of VSs from PaaS. RESTful interfaces are lightweight and can be useful in resource-constrained environment like vWSN.

The *second* requirement is that once created, the VS should be addressable similar to a VM in traditional IaaS. This will allow PaaS to seamlessly manage these VSs (e.g. start, stop, migrate and/or delete). Similarly, depending on the PaaS requirements and vWSN IaaS capabilities, certain parameters could be dynamically adjusted to configure VSs, such as sampling rate, reporting interval or even task migration (e.g. when monitoring dynamic events). In traditional IaaS, VMs get IP address and are accessible from anywhere, whereas the addressing mechanism of VSs depends on the platform and can be either a task-ID, MIDlet-ID, or some variation of 64bit IEEE hardware address. A mapping scheme at vWSN IaaS can be used to map global addresses to local ones.

TABLE I.  CONCEPTUAL DIFFERENCES BETWEEN VM AND VS

| Virtual Machine | Virtual Sensor |
|---|---|
| Logical representation of host machine | Logical representation of sensing capabilities of host sensor |
| Deployment of multiple OS-heterogeneous VMs | Middleware-dependent deployment of VS |
| Isolated deployment | Non-isolated deployment |
| Standard IP-based address mechanism | No standard mechanism to address |
| Unlimited power supply (for the physical host, i.e. server) | Battery operated (for physical host; i.e., sensor) |
| Proprietary and open source solutions | Currently no solutions |
| Uses resources of host machine (computing, storage) | Uses sensing capabilities of host sensor |

The *third* requirement is that the vWSN IaaS should be able to publish available services provided by the deployed sensors. For application development, PaaS will need to discover services provided by sensors, for example it might look for temperature service at a particular location for a certain duration and upon finding appropriate sensor, proceed to create a VS on it. In this situation a static or simple service description will not suffice for publication, instead it should include the spatial/temporal characteristics while considering the current load on that particular sensor. A centralized or distributed repository can be used for this purpose.

The *fourth* requirement is the lifecycle management and monitoring of VSs by the vWSN IaaS. In resource-constrained environments, VSs may not be as stable as VMs in traditional IaaS. Energy deficiency coupled with low bandwidth and hardware issues make it difficult to have always-on or always-available VSs. A robust VS lifecycle management and monitoring will be useful, e.g. in releasing VS (deleting them) when they are no longer in use, map application requirements from PaaS to available sensors, and help in fault detection and solution in the vWSN IaaS. However, satisfying spatial and temporal requirements is not trivial.

The *fifth and final* requirement is the support for intervWSN IaaS interactions. Typical WSN deployments will span over a geographic area and may need to interact with each other according to the requirements of the applications. Such interaction needs to involve SLAs, policy enforcement and of course deal with privacy and security issues.

Reference [8] provides an exhaustive survey of vWSN solutions but none of them meets all these requirements.

III. PROPOSED WSN IAAS ARCHITECTURE

In this section, we first present our previous vWSN architecture since we use it as a starting point for this work. Later we discuss our proposed vWSN IaaS architecture.

*A. vWSN Architecture*

This work is based on our previous work [10] in which we proposed a vWSN architecture shown in Fig. 1. It is a multilayer architecture that exploits the capabilities of individual sensor nodes to run concurrent application tasks at nodelevel and dynamically assembles such nodes at network-level for data sharing. The Physical layer has resource-constrained sensors (e.g., TelosB motes) and capable sensors (e.g., Java SunSpots). Since resource-constrained sensors may not support WSN network-level virtualization, they rely on capable Gateto-Overlay (GTO) nodes (e.g., base station nodes, sink nodes and capable sensors) for this purpose.

Next, in the Virtual Sensor layer, we have VSs that are abstractions of the application tasks run by the physical sensors. For each application, there is one VS running its task. The third layer is the Virtual Sensor Access layer. It consists of Sensor Agents (SAs) that provide platform independence by using standardized north-bound interfaces and proprietary south-bound interfaces. The final layer is the Overlay layer, which consists of multiple application overlays that use the deployed WSN. There are separate interfaces for data and control messages. The architecture is platform independent, applicable to different types of sensors, and does not cater any specific application domain.

*B. Proposed vWSN IaaS Architecture*

The proposed vWSN IaaS architecture is shown in Fig. 2. The following is the detailed description of the architecture.

The bottom two layers (WSN Infrastructure and Virtual Sensors) are similar to the ones in the previous architecture and consist of heterogeneous sensors, GTO nodes and virtual sensors. The functionality of these two layers and the roles of their entities are same as described in the Section III-A.

We have added a new layer called Virtual Sensor Manager, which contains two new functional entities: The VS Manager and VS Communicator. VS Manager receives requests to instantiate, start, stop, delete, and migrate VS. Tasks such as VS migration can only be accomplished if supported by the vWSN IaaS. The task code, which is to be run by the VS, is also disseminated through the VS Manager.

The VS Communicator supports platform-specific protocols to interact with different sensor platforms to promote platform heterogeneity. Examples of these protocols include IEEE 802.15.4, Bluetooth, Cellular, and RESTful.

Next, we renamed the Virtual Sensor Access layer from our previous architecture to the Virtualized WSN Infrastructure Management layer to make it more appropriate for this work. It now contains several new entities in addition as well as SAs. SAs interact with the WSN PaaS components on behalf of the VS in order to provide platform independence. The additional entities are described as follows.

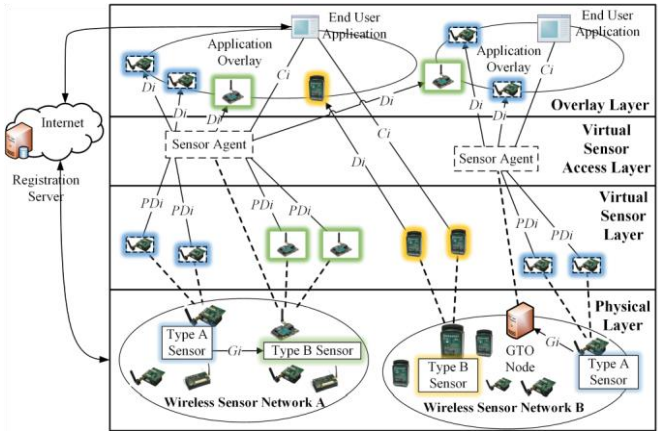

Fig. 1. Original vWSN Architecture

The Sensor Description Repository contains all relevant information about the deployed sensors, including their type, properties (i.e., protocol, data format, supported sampling intervals, physical location and supported units) and capabilities (i.e., sensing abilities). The repository can be distributed or centrally located and it is the responsibility of the WSN infrastructure owner to keep it up-to-date.

The Sensor Discovery entity, interacts with the repository to search for the required sensors using any criteria, e.g., sensor type, location and its availability.

The VS Provider is the main entity that receives VS creation requests from the WSN PaaS. Based on these requests, sensors are selected from the repository. The VS Provider also makes decision about when to create, start, or stop a VS by communicating with the VS Manager. There is also a small cache of the most recent sensors used by applications to prevent the need to search for sensors every time a request comes from the PaaS.

The VS Configurator entity prepares task codes based on the requests received from the VS Provider. These tasks will be run by a given VS. VS Configurator uses platform-specific code templates that allow for configurable parameters. A code template is a skeleton code file that does nothing useful on its own but can read from a parameter list and run a desired task. An example is the skeleton code that reads a manifest file (i.e., used in Java SunSpot platform) to initialize parameters such as sensor type, sampling interval, desired unit, and an end-point address to send data output. Creating these manifest files on the fly is programmatically simple and can be easily achieved.

The VS Configurator should ideally be implemented in a modular fashion to allow for the possibility of adding future code templates when new types of sensors are deployed. Additionally, VS Configurator compiles and generates the final executable code (e.g., jar file for Java SunSpot).

The role of VS Scheduler entity is to create, start, stop, and disseminate task codes either right away or at a later time, depending on the application requirements. It interacts with VS Manager to accomplish this.

The final layer is the Cloud Management layer, which includes an entity called the IaaS Access/Control Interface.

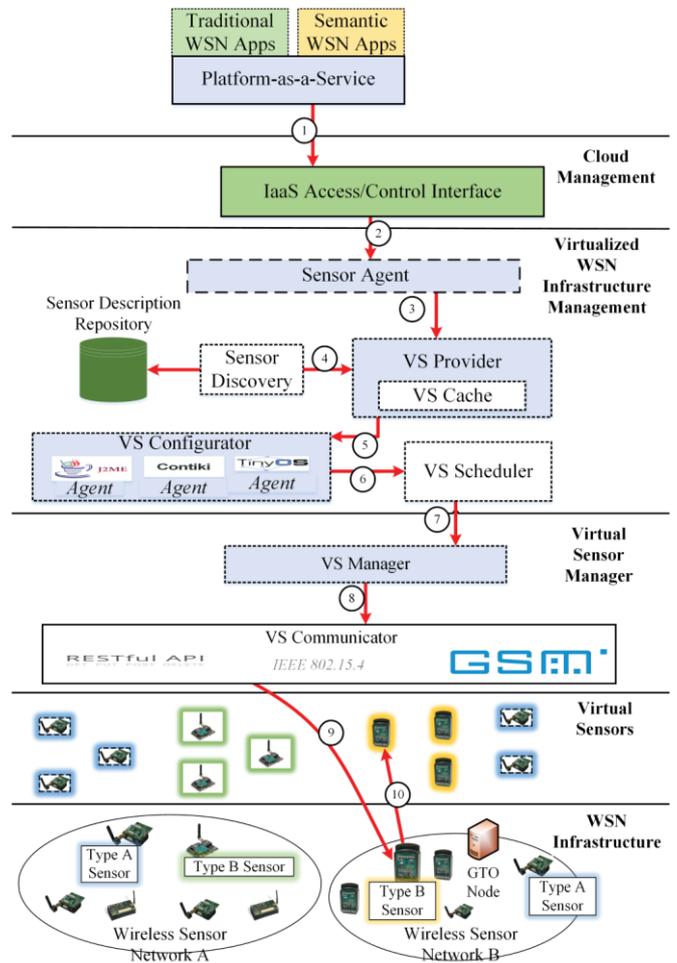

Fig. 2. Proposed vWSN IaaS Architecture

This interface exposes a RESTful API that allows multiple users (i.e. PaaS) to interact with the deployed vWSN IaaS through a set of REST-based operations.

IV. AN EARLY IMPLEMENTATION AND RESULTS

In this section, we first discuss a simple scenario used for implementation. Then, we present our implementation choices and prototype setup. Next, we discuss performance metrics and finish off this section with a discussion on the results.

*A. Implementation Scenario*

A smart home application is required that allows home owners to configure the use of their appliances when environmental conditions change. For example, the A/C should start automatically when temperature exceeds a given threshold. Similarly, the deck lights should be turned-on automatically when natural light drops below a given threshold.

The developer first discovers the light and temperature services to design and create the smart home application. When the application is deployed, the PaaS allocates an application container along with two REST-Based interfaces. One interface is for the VS corresponding to the light sensor and the other for the VS corresponding to the temperature sensor.

*B. Implementation Choices and Prototype Setup*

The WSN infrastructure consists of Java SunSpots, which have multiple on-board sensing capabilities. Unlike the earlier generation of sensor nodes, Java SunSpots are quite capable and are based on Java 2 Micro Edition (J2ME), which makes them easier to program. The Squawk VM supports multithreading, making them suitable for our work. We used two Java SunSpot kits: two base station nodes and four SunSpots with on-board sensors. The vWSN IaaS layers were implemented as a standalone application.

We programmed a simple PaaS, as a standalone Java application. Eclipse IDE and JDK 1.7 were used for the application development. The application code was annotated with a description of the VS services and was given to the developers beforehand. The smart home application was developed as a simple Java application.

We used two laptops for the prototype. The first one had the PaaS, and the second one had the vWSN IaaS. The two laptops were connected via Ethernet and established as a LAN network. The vWSN IaaS laptop was connected to the Java SunSpot base stations to communicate with the remote SunSpots Over-the-Air (OTA).

*C. Performance Metrics*

The performance of the prototype was assessed in terms of the following metrics: VS Creation Delay (VSCD) and VS Start Time (VSST). The time spent between the moment the developer sends the application code to the PaaS for deployment and the moment the PaaS sends the creation requests to the vWSN IaaS was found to be negligible.

VSCD is the time spent between the moment the WSN infrastructure receives the VS creation request from the PaaS and the moment the VS is successfully created. Because it is required to create a shared base station instance to communicate with remote Java SunSpot OTA, we measured two types of VSCD. In the first type, the shared base station instance is created once and used repeatedly for VS creation, hence it only shows VS creation delay. In the second type, a shared base station is created every time a VS creation request is received from the PaaS, hence it shows VS creation delay plus the delay to create the shared base station instance.

VSST is the time spent when the WSN infrastructure receives the VS start request from the PaaS and when the corresponding VS is successfully started. All experiments were repeated 50 times with a confidence interval of 95%.

*D. Results*

Fig. 3 shows the values of both types of VSCD over 50 iterations. On average, it took about 14.973 seconds to create a VS on a remote Java SunSpot when the shared base station was created once. However, for the second type of VSCD, the average value increased by around 62%, to 24.282 seconds. One reason for this increase is that the shared base station instance spends some time probing for the available remote SunSpots. This delay is unavoidable and is not related to our architecture. Another reason for both of these high values is the fact that we used Ant build tool (as required by Java SunSpot platform) to first build, compile and create the executable file and then send it to remote Java SunSpots OTA. The last step included the delay to synchronize the target Java SunSpot. The actual dissemination of the VS code to the remote SunSpot took the very less time.

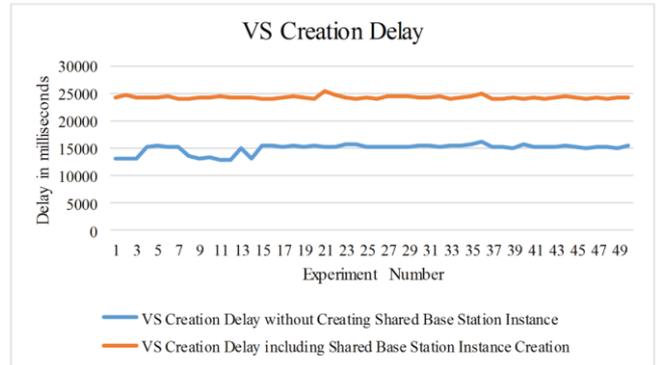

Fig. 3.  VS Creation Delay

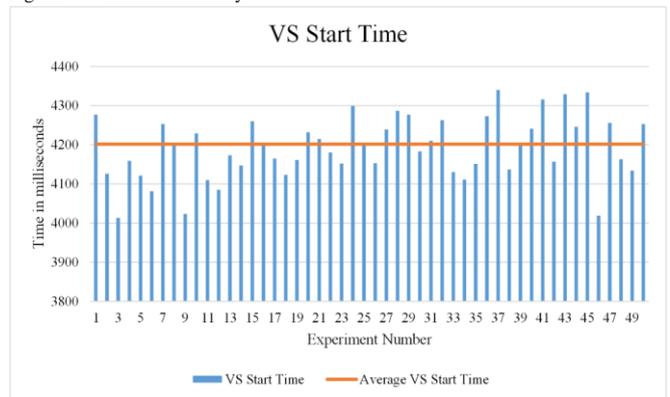

Fig. 4.  VS Start Time

Fig. 4 shows the VSST of the 50 iterations. On average, it took 4.2 seconds to start the newly created VS after receiving the request from the PaaS. Again, this delay included the remote SunSpot synchronization delay before the newly created VS was started. Overall, these results are promising and prompt us to explore the problem area further in order to provide more optimized solutions.

V. LESSONS LEARNED AND FUTURE WORK

In this work, we have learned several lessons and have also identified many research issues to further pursue.

The *first* lesson learned is that while RESTful interfaces provide an easy way to access VSs, however, integrating them with existing open source PaaS (e.g., CloudFoundry) will be quite challenging. The *second* lesson is that there are other capable sensor kits in addition to Java SunSpots, such as Preon32 sensor kits from Virtenio GmbH [26] (Java-based and similar to SunSpots) and Phidgets kit [27]. The *third* lesson learned is that during the creation of VS on a Java SunSpot, the execution of existing VSs is not disturbed. This feature is very useful for ensuring that existing applications do not suffer when new ones utilize a SunSpot. Similarly, the VS migration feature is also supported by SunSpots, and we intend to work on this in the future. The *fourth* lesson is that the delay associated the creation of VSs will largely depend on the platform. Java SunSpots need Ant build tool whose performance heavily depends on the installed Java version and the workload on the host machine.

As for the future work, *first* we plan to work towards the complete implementation of the architecture as presented in Section III-B and satisfy all the requirements mentioned in Section II-C. To this end, we intend to incorporate additional sensor platforms to allow for the heterogeneity of sensor nodes. The Preon32 and Phidgets kits are two possible candidates. *Second* we plan to work on exploiting the capabilities of available Java SunSpot kits by implementing the full features (e.g., VS stop, delete and migration to another remote SunSpot on the fly) they offer. *Third* we plan to provide the VS reservation mechanism by implementing a VS Scheduler entity, which would be very useful for a business model wherein a vWSN IaaS could be leased to users against certain incentives [28].

While this work focuses on the vWSN IaaS, we also felt the need to have a capable PaaS for vWSNs IaaS, because existing PaaS solutions do not consider the possibility of using VSs for application and service provisioning. For example, there is a need to discover and manage VSs and their details at the PaaS level but currently there is no solution for this. Instead most solutions simply receive sensor data and use it without taking full advantage of a vWSN IaaS.

## VI. CONCLUSION

In this paper we have presented an architecture for a competent vWSN IaaS that is able to interact with the PaaS to support the concurrent VS-based applications and services deployment on-demand. The architecture uses the principles of cloud computing and the basics of WSN virtualization to offer WSN deployments as IaaS. Using a capable sensor kit, an early implementation has demonstrated its feasibility. We have also identified several interesting and potent research issues and plan to tackle them in future contributions.

## ACKNOWLEDGMENT

This work was supported in part by the Natural Science and Engineering Council of Canada (NSERC) Canada Research Chair in End-User Service Engineering for Communications Networks and by an NSERC Discovery Grant.